\documentclass[pra,aps,onecolumn,showpacs,epsfig]{revtex4}
\usepackage{amssymb}
\usepackage{graphicx}% Include figure files
\usepackage{dcolumn}% Align table columns on decimal point
\usepackage{bm}% bold math
\usepackage{epsfig}
\newcommand{\beq}{\begin{equation}}
\newcommand{\eeq}{\end{equation}}
\newcommand{\beqa}{\begin{eqnarray}}
\newcommand{\eeqa}{\end{eqnarray}}
\newcommand{\ba}{\begin{array}}
\newcommand{\ea}{\end{array}}
\usepackage{color}
\begin{document}

\title[Quantum interferometry at zero and at finite temperature with two-mode bosonic Josephson junctions]
{Quantum interferometry at zero and finite temperature with two-mode bosonic Josephson junctions}
\author{G. Mazzarella}
\address{Dipartimento di Fisica e Astronomia ``Galileo Galilei''
and Consorzio Nazionale Interuniversitario per la Scienze Fisiche della
Materia (CNISM), Universit$\grave{a}$ degli Studi di Padova,
Via Marzolo 8, I-35131 Padova, Italy}
\date{\today}
\begin{abstract}
We analyze phase interferometry realized with a bosonic Josephson junction
made of trapped dilute and ultracold atoms. By using a suitable phase
sensitivity indicator we study the zero temperature junction states {\it useful} to achieve sub shot-noise precisions.
Sub shot-noise phase shift sensitivities can be reached even at finite temperature under a suitable choice of the junction state. We infer a scaling law in terms of the size system (that is, the number of particles) for the temperature at which the shot-noise limit is not overcome anymore.

\end{abstract}
\maketitle

\section{Introduction}
The quantum interferometry with $N$ particles has attracted much attention due
to the possibility to achieve phase sensitivities below the shot-noise limit
given by $1/\sqrt{N}$, that is the limit obtained by using classical states \cite{caves1}.
To overcome this limit, genuine quantum effects, such as the squeezing and the entanglement, are strictly necessary \cite{winekit,darphil,lorenzo,giovannetti}.

The theoretical study of this subject is at the base of various proof-of-principle experiments - with fixed number of photons \cite{photons} or ions \cite{ions} - reaching a sub shot-noise precision, and of many applications ranging from quantum lithography to quantum positioning and clock synchronization \cite{giovannetti}. Linear interferometers are crucial systems to realize optimal input states \cite{caves1,cavesetal}, perform adaptive phase measurements schemes \cite{adaptive}, and understand the influence of particle losses \cite{losses}.

A topic of great interest is the quantum metrology with two-mode bosonic Josephson junctions (BJJs) made of trapped ultracold and dilute atoms \cite{bec1,bec2,winekit,hyllus,grond1,grond2,pezzelast,pezze2,anna2}. The two modes
can be realized, for example, by using a far off-resonance laser barrier to split a trapped Bose-Einstein condensate
in two parts, left and right. The microscopic dynamics of such a system is efficiently described
by the two-site Bose-Hubbard (BH) Hamiltonian \cite{milburn}.
By diagonalizing this Hamiltonian, one knows the junction states that can be used as input of an atomic interferometer. The atomic interferometry relies on the accumulation
of a phase shift $\theta$ between the two modes during a given time of evolution.
%Then, a recombination of the splitting - realized, for example, by a non-adiabatically
%fast extinction of the double-well potential \cite{schumm} - allows to read out the interference
%fringes, from which $\theta$ is inferred. 
The corresponding interferometric sequence can be described by matrices of rotation (of an angle $\theta$ about a
given axis on the Bloch sphere) acting on the input state \cite{pezze,anna2}.
As observed in \cite{huang2,ferriniphd}, these rotations are implementable since the paramateres of the BH
Hamiltonian -  i.e. the on-site interatomic interaction $U$ and the hopping $J$ between the two wells - are experimentally tunable. For rotations of an angle $\theta$ about the $x$ axis, one lets act on the input state the temporal evolution operator associated with the
Hamiltonian $\hat{H}_x=-2J \hat{S}_x$ ($\hat{S}_x$ is the component along $x$ axis on the Bloch sphere of
the pseudospin operator in the Schwinger representation \cite{kim}).
The Hamiltonian $\hat{H}_x$ can be realized, from the initial two-site
BH Hamiltonian, e.g. by switching off $U$ via the Feshbach resonance technique. The phase shift $\theta$ is got at the time $t=\hbar\theta/2J$. Rotations of
the same amount about the $z$ axis can be obtained by the temporal evolution operator associated with the
Hamiltonian $\hat{H}_z=-\delta E \hat{S}_{z}$ ($\hat{S}_z$ is the component along the $z$ axis on the Bloch sphere of the
pseudospin operator) on the input state, with $\delta E$ the difference between the energies of the two wells \cite{anna1}.
The Hamiltonian $\hat{H}_z$ can be realized by switching off both the interatomic
interaction and the hopping amplitude, for instance, by rising the central
barrier. One therefore gets the phase shift $\theta$ after a time
$t=\hbar\theta/\delta E$. The rotations about the $y$ axis can be realized by
suitable combinations of rotations about the $x$ and $z$ axes.

The central issue of the interferometry is the phase shift
estimation. To face this problem, one has to calculate the quantum Fisher information (QFI) \cite{lorenzo,wooters,helstrom,braunstein}.
The QFI is related to the bound on the precision with which $\theta$ can be determined.
For a given input state, the best achievable precision is the reciprocal of the squared
root of QFI, known as the quantum Cram\'er-Rao (QCR) bound \cite{pezze2,helstrom}.

The authors of \cite{pezze} consider the squared root of the ratio between $N$ and the QFI. This quantity,
denoted by $\chi$, can be calculated for each rotation axis both at zero (pure
states) and at finite temperature (mixed states). The states of a given $N$-particle system are {\it useful}
for achieving sub shot-noise precision if $\chi<1$ which represents a sufficient condition for multiparticle entanglement \cite{lorenzo}. Such a condition ensures that QCR bound is smaller than $1/\sqrt{N}$. The question whether or not all pure entangled $N$-particle states are useful
for sub shot-noise metrology was considered in \cite{hyllus}, but the atom-atom
interaction does not explicitly appear. The impact of this interaction on the atomic interferometry
is taken into account in \cite{grond1,grond2} for repulsive bosons. In particular, Ref.
\cite{grond2} deals with the temperature effects on systems used for two-mode interferometry. As far we know, a systematic study of the robustness of the condition
$\chi <1$ against the temperature lacks.

In this work, we consider a bosonic Josephson junction realized with $N$ atoms
confined by a symmetric one-dimensional double-well potential
within the two-site Bose-Hubbard model framework. To analyze the usefulness of
the BJJ states for quantum interferometry, we diagonalize the BH Hamiltonian and calculate $\chi$ - named, in this paper,
phase sensitivity indicator (PSI) -  both at zero and at finite temperature. At
zero temperature, we study the PSI as a function of the boson-boson interaction
from the strongly attractive regime to the repulsive one. In this way we explore a wide range of ground-states sustained by the BH Hamiltonian by extending the study presented in \cite{grond1} to attractive interactions and
continuing the study presented in \cite{hyllus} making explicit the role played by the atom-atom interaction.
For attractive bosons the BJJ ground-states are useful input states only for the interferometry
based on rotations about the $z$ axis. When the bosons are repulsively interacting, a range of interaction exists
over which it is possible to achieve sub shot-noise precisions only
for rotations about $y$ axis-based interferometry. If the boson-boson repulsion becomes sufficiently strong, the shot-noise limit is overcome also for metrology realized via rotations about the $x$ axis.

In the presence of the temperature, we analyze the thermal effects on the quantum
metrology usefulness of the junction states. We continue the
analysis of \cite{hyllus} by considering mixed states and further develop the issue addressed in \cite{grond2}
by analyzing the phase sensitivity indicators as functions of the temperature
for different interaction strengths. The interesting result that we find is that, over a certain range of temperatures,
the junction states are useful input states for quantum interferometry
based on rotations about the $y$ axis in the repulsive regime, and about the $z$ axis
for attractive bosons: in the former case, an increase of the interaction
between the bosons enhances the robustness of the condition $\chi <1$ against the
temperature, while in the latter one this enhancement is observed by lowering the
absolute value of the interaction strength. When the temperature becomes sufficiently high the possibility to
reach sub shot-noise precisions is destroyed. We evaluate the temperature at which occurs the breakdown of the quantum
metrology usefulness. For repulsive bosons, we analyze the size effects on the robustness of the quantum metrology usefulness condition by obtaining a scaling law with the number of bosons for the breakdown temperature. We point out that to an increasing the number of particles a rising of this breakdown temperature corresponds.

\section{The model Hamiltonian}

We consider $N$ identical bosons of mass $m$ confined
by a trapping potential $V_{trap}({\bf r})$. We suppose that this
potential is realized by superimposing to an isotropic harmonic
confinement in the the transverse radial plane
a double-well potential (DWP) $V_{DW}(x)$ in the axial direction $x$.
Then,
\beq
\label{trap}
V_{trap}({\bf r})=V_{DW}(x) +\frac{m\,\omega_{\bot}^2}{2}\,(y^2+z^2)
\;,\eeq
where $\omega_{\bot}$ is the trapping frequency in the radial plane.
We suppose that the system is quasi one-dimensional (1D) due to a strong transverse radial harmonic confinement. In particular, the transverse energy $\hbar \omega_{\bot}$ is much larger than
the characteristic trapping energy of bosons in the axial direction. If the two wells are symmetric, the effective two-sites Bose-Hubbard (BH) Hamiltonian \cite{milburn}
\beq
\hat{H} = -J\big(\hat{a}^{\dagger}_L\hat{a}_R
+\hat{a}^{\dagger}_R\hat{a}_L\big)
+\frac{U}{2}\big( \hat{n}_L (\hat{n}_L -1)
+ \hat{n}_R (\hat{n}_R -1)\big) \;
\label{twomode}
\eeq
describes the microscopic dynamics of the system. In the Hamiltonian
(\ref{twomode}) $\hat{a}_{k},\hat{a}^{\dagger}_{k}$ ($k=L,R$)
are bosonic operators satisfying the algebra
$[\hat{a}_{k},\hat{a}^{\dagger}_{l}]=\delta_{kl}$;
$\hat{n}_{k}=\hat{a}^{\dagger}_{k}\hat{a}_{k}$ is the number of
particles in the $k$th well; $U$
is the boson-boson interaction amplitude; $J$ is the tunneling matrix
element between the two wells.
%We, moreover, observe that the total number operator
%\beq
%\hat{N}= \hat{n}_{L} +\hat{n}_{R}
%\eeq
%commutes with the Hamiltonian (\ref{twomode}).
In the following we shall analyze the system both at zero and at finite
temperature. We assume that $k_BT$ - $k_B$ is Boltzmann's constant - is not much larger than the
gap of the lowest doublet, while it is much smaller than the gap
between the first and the second doublet of the DWP linear problem. In this
situation, the lowest doublet will be the only one to be occupied. The spectrum of the Hamiltonian (\ref{twomode}) is determined by solving the eigenproblem
\beq
{\hat H} |E_j\rangle = E_j |E_j \rangle
\eeq
for a fixed number $N$ of bosons. In this case
the Hamiltonian ${\hat H}$ can be represented by
a $(N+1)\times (N+1)$ matrix in the Fock basis $|i,N-i\rangle$ (the
left (right) index denotes the number of bosons in the left (right) well), with $i=0,...,N$, and
$j=0,...,N$ denoting the $j$th excited state.

The macroscopic parameters in the Hamiltonian (\ref{twomode}) are explicitly related to the atom-atom coupling
constant $g=4\pi\hbar^2a_s/m$ - $a_s$ being the s-wave scattering
length - and to the other microscopic parameters, i.e. the atomic mass $m$ and the frequency $\omega_{\bot}$
of the harmonic trap (see, for example, \cite{ajj2}).
The on-site interaction amplitude $U$ is positive (negative) if $a_s$ is
positive (negative), so that it may be changed at will by Feshbach resonance.
Note that when $a_s<0$, to avoid the collapse, the system has to be
prepared in such a way that the $\displaystyle{|U| \lesssim \frac{0.4}{N}\hbar (\omega_{x}\omega_{\bot})^{1/2}} $, \cite{gammal}, with $\omega_{x}$ the trapping frequency of the single well.

The hopping amplitude is given by
\beq
\label{hopping}
J=\frac{\epsilon_1-\epsilon_0}{2}
\;,
\eeq
where $\epsilon_0$ and $\epsilon_1$ are the ground-state and the first excited
state energies of a single boson in the double-well \cite{fermidyn}.

Notice that at fixed number $N$ of bosons, the ground-state of the system is controlled by the dimensionless
parameter $\zeta=U/J$ \cite{cats}. This parameter shall be used also throughout the present work to discuss the role of the boson-boson interaction in determining the system properties. In terms of $\zeta$ the previous condition about the collapse thus reads $\displaystyle{|\zeta| \lesssim \frac{0.4}{N J} \hbar(\omega_{x}\omega_{\bot})^{1/2}}$. Then, fixed $N$ and the radial frequency, the collapse of the bosonic cloud can be prevented by suitably adjusting the parameters of the DWP.

\section{Analysis}
We focus on the following problem: given a state described by the density matrix $\hat{\rho}$, is this
an useful input state for quantum interferometry?

As commented in the introduction, a quantum interferometer can be described by rotation of the input state $\hat{\rho}$ of an angle $\theta$ about the $\gamma$ axis ($\gamma=x,y,z$) on the Bloch sphere. Then, the output state $\hat{\rho}_{out}$ will be given by
\beq 
\label{rotation}
\hat{\rho}_{out}(\theta)=\exp(i \theta\hat{S}_{\gamma}) \hat{\rho} \exp(-i \theta \hat{S}_{\gamma})
\;.\eeq
The operators $\hat{S}_{\gamma}$ are the
pseudospin operators in the Scwhinger boson representation \cite{kim}, i.e.
\beqa
\label{pseudospin}
&&\hat{S}_{x}=\frac{1}{2}\,(\hat{a}^{\dagger}_L\hat{a}_R
+\hat{a}^{\dagger}_R\hat{a}_L)\nonumber\\
&&\hat{S}_{y}=-\frac{i}{2}\,(\hat{a}^{\dagger}_L\hat{a}_R
-\hat{a}^{\dagger}_R\hat{a}_L)\nonumber\\
&&\hat{S}_{z}=\frac{1}{2}\,(\hat{n}_L-\hat{n}_R) \;.
\eeqa
%As observed in \cite{huang2,ferriniphd}, the rotations
%(\ref{rotation}) can be implemented since both the interatomic interaction $U$
%and the hopping $J$ are experimentally tunable parameters. In the case of
%a rotation of an angle $\theta$ about the $x$ axis, one lets that
%the temporal evolution operator associated with the
%Hamiltonian $\hat{H}_x=-2J \hat{S}_x$ acts on the state $\hat{\rho}$.
%The Hamiltonian $\hat{H}_x$ can be realized, from the
%Hamiltonian (\ref{twomode}), e.g. by switching off the interatomic interaction
%via the Feshbach resonance technique. The phase shift $\theta$ is got at the time $t=\hbar\theta/2J$. A rotation of
%the same amount about the $z$ axis can be obtained
%by means of the action of the temporal evolution operator associated with the
%Hamiltonian $\hat{H}_z=-\delta E \hat{S}_{z}$ on $\hat{\rho}$, where $\delta E$
%is the difference between the energies of the two wells \cite{anna1}.
%The Hamiltonian $\hat{H}_z$ can be realized by introducing
%- in the initial system described by the Hamiltonian (\ref{twomode}) -
%an asymmetry between the two wells and by switching off both the interatomic
%interaction and the hopping amplitude, e.g. by rising the central
%barrier. In this setup, one gets the phase shift $\theta$ after a time
%$t=\hbar\theta/\delta E$. The rotations about the $y$ axis can be realized by combinations of
%rotations about the $x$ and $z$ axes: $e^{i\theta \hat{S}_y}=e^{i
%\pi/2 \hat{S}_x} e^{i\theta \hat{S}_z} e^{-i\pi/2 \hat{S}_x}$.

\subsection{Zero-temperature analysis}
At zero temperature the system is in a pure state which depends on the
dimensionless interaction parameter $\zeta=U/J$ \cite{cats}.
In the deep attractive regime, $|U| \gg J$, the ground-state evolves towards the following macroscopic superposition state
\beq
\label{cat}
|CAT \rangle=\frac{1}{\sqrt{2}}(|N,0\rangle+|0,N\rangle)
\;.\eeq
On the other hand, when the repulsion between the bosons is very
strong, $U \gg J$, the ground-state is close to the twin-Fock state
\beq
\label{twin}
|FOCK\rangle=|\frac{N}{2},\frac{N}{2}\rangle
\;.\eeq

It is well known that the states $|CAT \rangle$ and $|FOCK\rangle$ are useful for interferometry, respectively, based on rotations about $z$ and $y$ axes \cite{giovannetti,pezze-smerzi,hb}. Since the boson-boson interaction can be tuned by Feshbach resonance, understanding what happens, point by point, in a sufficiently wide range of $\zeta$ is physically meaningful. The density matrix is $\hat{\rho}=|\psi \rangle \langle \psi|$. By following \cite{hyllus},
we consider the question whether pure states are useful or not for quantum interferometry. We calculate the phase estimation indicator (PSI) for a pure state $|\psi \rangle$, say $\chi_{ps,\gamma}$, which is given by \cite{pezze}
\beq
\label{chips}
\chi_{\gamma,ps}=\sqrt{\frac{N}{F_{ps,\gamma}}}
\;.\eeq
Here $F_{ps,\gamma}$ is the pure state quantum Fisher information \cite{pezze,braunstein}
\beq
\label{purefisher}
F_{ps,\gamma}=4\langle \psi|(\Delta \hat{S}_{\gamma})^2|\psi \rangle \;
\eeq
with $\Delta$ the variance of the pseudospin operators $\hat{S}_{\gamma}$ (\ref{pseudospin}).
%In terms of the coefficients of the expansion (\ref{eigenstate}), the quantum
%Fisher informations $F_{ps,x}$, $F_{ps,y}$, and $F_{ps,x}$, are, respectively
%\beqa
%\label{fx}
%F_{ps,x} & = &
%\sum_{i=0}^{N}
%c_{i}^{(0)} c_{i+2}^{(0)}
%\sqrt{(i+1)(i+2)(N-i)(N-i-1)} \nonumber\\
%&+&c_{i}^{(0)}c_{i-2}^{(0)}\sqrt{(i)(i-1)(N-i+1)(N-i+2)}\nonumber\\
%&+&c_{i}^{(0)})^2\,(N+2i(N-i))-\nonumber\\
%&&4\,\big(\sum_{i=0}^{N} c_{i}^{(0)}c_{i+1}^{(0)} \sqrt{(i+1)(N-i)}\big)^2
%\;,\eeqa

%\beqa
%\label{fy}
%F_{ps,y} & = &
%-\sum_{i=0}^{N}
%c_{i}^{(0)}c_{i+2}^{(0)}\sqrt{(i+1)(i+2)(N-i)(N-i-1)} \nonumber\\
%&+&c_{i}^{(0)}c_{i-2}^{(0)}\sqrt{i(i-1)(N-i+1)(N-i+2)}\nonumber\\
%&-&(c_{i}^{(j)})^2\,(N+2i(N-i)) \;,
%\eeqa

%\beq
%\label{fz}
%F_{ps,z}=\sum_{i=0}^{N}(c_{i}^{(0)})^2\,(2i-N)^2\;.
%\eeq
It can be shown that for any separable input state of $N$ particles in two modes ($N$ qubits) one has
$F_{ps,\gamma} \le N$ \cite{lorenzo}. Therefore the condition $F_{ps,\gamma} > N$, i.e.
\beq
\label{sc}
\chi_{\gamma,ps} <1
\eeq
is a sufficient condition for $\hat{\rho}$ to be entangled \cite{lorenzo}.
It is very important, now, to stress that the condition (\ref{sc}) is related to the multiparticle entanglement, i.e. to the indistinguishability of the bosons \cite{pezze}. From this perspective, states separable according to the partition left-right point of view (i.e., by calculating the entanglement entropy $S$ \cite{bwae}) but multiparticle entangled exist. In fact, for the twin Fock-state (\ref{twin}) $S$ is zero, but $\chi_{y,ps} <1$.

For an arbitrary interferometer and phase estimation strategy, the phase
precision for a given input state $\hat{\rho}$ is limited by
the quantum Cram\'er-Rao (QCR) bound - see \cite{braunstein} - given by
\beq
\label{qcr}
\Delta \theta_{QCR}=\frac{1}{\sqrt{F_{ps,\gamma}}}=\frac{\chi_{\gamma,ps}}{\sqrt{N}}
\;.\eeq
We thus see that the condition (\ref{sc}) provides the class of entangled states for which
$\Delta \theta_{QCR}<1/\sqrt{N}$, so that the shot-noise limit is overcome. The use of
such entangled states as input of an interferometer - realizing the unitary transformation
$\exp(-i \theta \hat{S}_{\gamma})$ - makes possible a phase estimation with a precision
higher than any interferometer using classical (separable) states. On the other
hand, the class of entangled states for which $\chi \ge 1$ cannot provide sensitivities
higher than the classical shot noise \cite{pezze}.
\begin{figure}[ht]
\epsfig{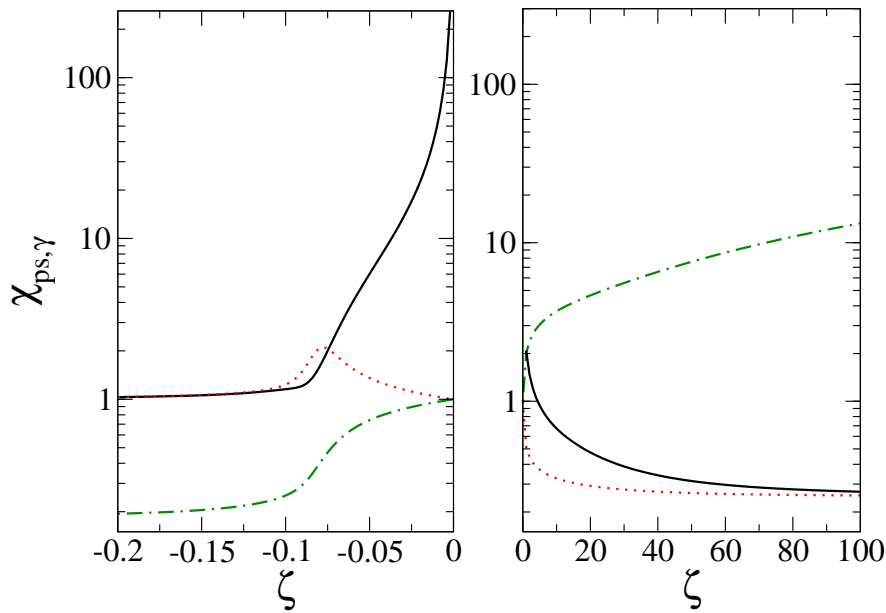}
\caption{(Color online). The phase sensitivity indicator  $\chi_{ps,\gamma}$ as a function of the dimensionless parameter $\zeta=U/J$ at $T=0$. Left panel: attractive bosons, $\zeta<0$. Right panel: repulsive
bosons, $\zeta>0$. Solid line: $\chi_{ps,x}$; dotted line:  $\chi_{ps,y}$; dot-dashed line:
$\chi_{ps,z}$. Number of bosons $N=30$. Notice that the vertical axis of
both the panels is in logarithmic scale.}
\label{chipure}
\end{figure}
We calculate $\chi_{\gamma,ps}$  by using Eq. (\ref{purefisher})
in Eq. (\ref{chips}) for each $\gamma$. The behavior of the pure state PSI
as a function of $\zeta=U/J$ is shown in Fig. \ref{chipure}. In the attractive regime, when one increases $|\zeta|$ starting from zero, the ground-state of the Hamiltonian (\ref{twomode}) evolves from an atomic coherent state to an entangled superposition of macroscopic states, i.e $|CAT \rangle$, \cite{cats,dellanna}. In these papers, it was observed that by rising the boson-boson attraction, the system exhibits a dephasing in terms of losing of inter-well coherence characterized by $2|\langle \hat{a}^{\dagger}_R\hat{a}_L\rangle|/N$, \cite{cats,dellanna}, to which $\hat{S}_x$ and $\hat{S}_y$ are related, see Eq. (\ref{pseudospin}). We, then, expect that the interferometry candidate to achieve sub-shot noise precisions will be based on rotations about the $z$ axis. From the left panel of Fig. \ref{chipure}, it can be seen that $\chi_{ps,x}$ (solid line) and $\chi_{ps,y}$ (dotted line) are, always, greater than one. The quantity $\chi_{ps,z}$ (dot-dashed line) is always smaller than one except for $\zeta=0$ when $\chi_{ps,z}=1$.

Let us focus, now, on the repulsive regime. From the right panel of Fig. \ref{chipure}, it can be observed that when the repulsion is strong enough, the quantum metrology usefulness condition is met for the unitary transformation $exp(-i\theta\hat{S}_{x})$. The BJJ ground-states achievable with repulsive interactions allow for reaching sub-shot-precisions for interferometry in the $y$ direction except that when the interaction is absent. When $\zeta >0$, we have that $\chi_{ps,z} \ge 1$.

So far we have taken into account bosons at zero temperature. In realistic
experiments, however, quantum systems work at finite temperature. Since for low
dimensional geometries the effects due to thermal fluctuations are quite important
\cite{ft}, it is mandatory focus on the influence of the temperature in achieving
sub-shot noise precisions.

\subsection{Finite temperature analysis}
At finite temperature the system is in a statistical mixture of
states. The mixed state is described by the density matrix $\hat{\rho}$ which, in
the canonical ensemble, reads
\beq
\label{mixed}
\hat{\rho}=\sum_{j=0}^{N} \frac{e^{-\beta E_j}}{Z} |E_j\rangle\langle E_j|\;
\eeq
with $Z$ the partition function given by $Z=\sum_{j=0}^{N}\ e^{-\beta E_j}$, where $\beta=1/k_BT$, $k_B$ the constant of Boltzmann and $T$ the absolute
temperature. For mixed states, the quantum Fisher information, $F_{QT,\gamma}$, is given by \cite{braunstein}
\beqa
\label{pezze}
F_{QT,\gamma} &=&
\sum_{j,k=0}^{N}\frac{2\,(e^{-\beta E_j} -e^{-\beta E_k})^2}{Z\,(e^{-\beta E_j}
+e^{-\beta E_k})}|\langle E_{j}|\hat{S}_\gamma|
E_{k}\rangle|^{2} \;.\nonumber\\
\eeqa
Also for mixed states (\ref{mixed}), it is possible to define a PSI,
$\chi_{\gamma}$, \cite{pezze}
\beq
\label{chi}
\chi_{\gamma}=\sqrt{\frac{N}{F_{QT,\gamma}}}
\;.\eeq

\begin{figure}[ht]
\epsfig{file=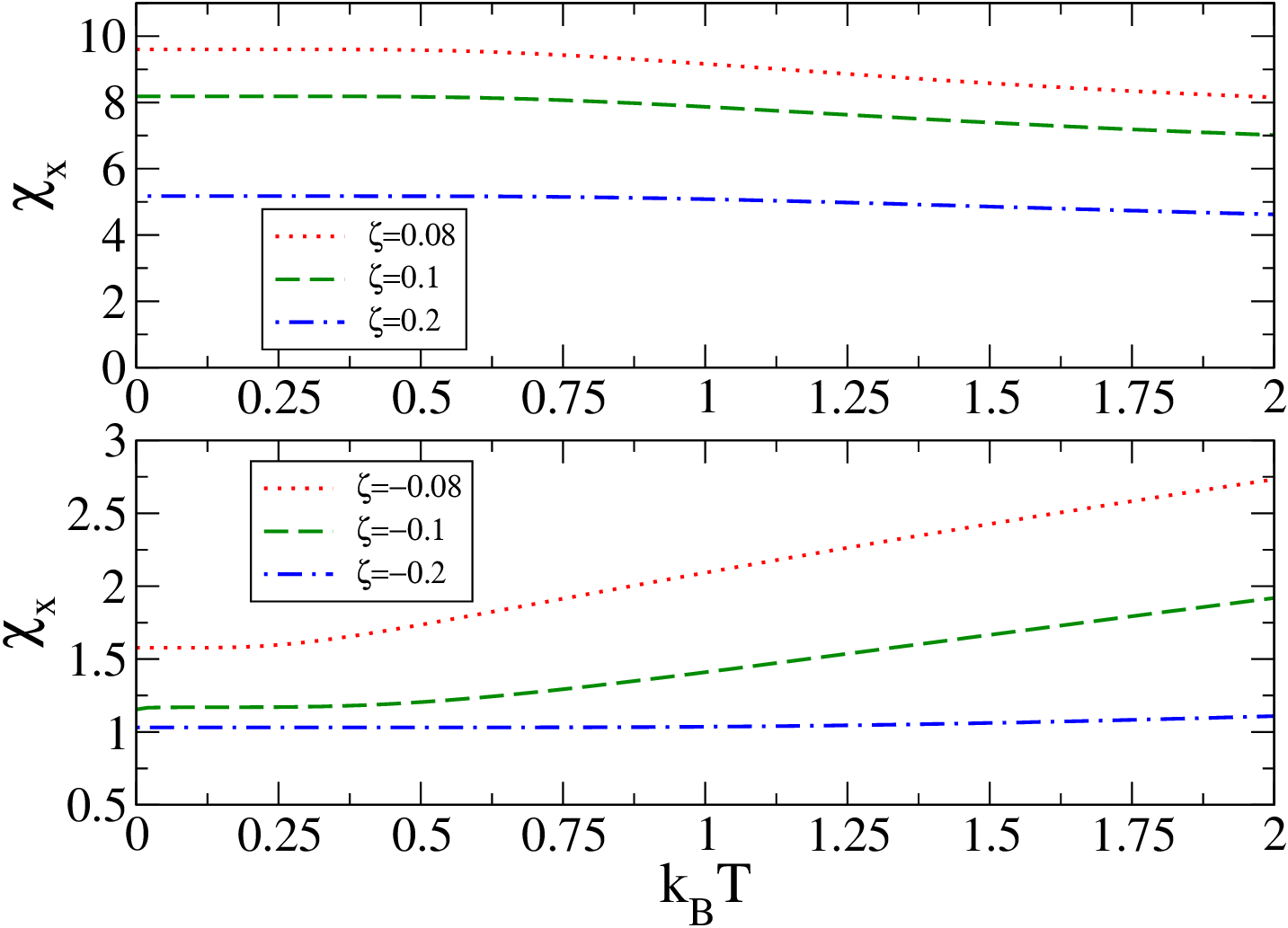,width=0.65\linewidth,clip=}
\caption{(Color online). The phase sensitivity indicator $\chi_{x}$ vs $k_B
T$. Upper panel: repulsive bosons, $\zeta >0$. Lower panel: attractive bosons, $\zeta<0$. Number of bosons $N=30$.  $k_BT$ is in units of $J$.}
\label{fig6}
\end{figure}

\begin{figure}[ht]
\epsfig{file=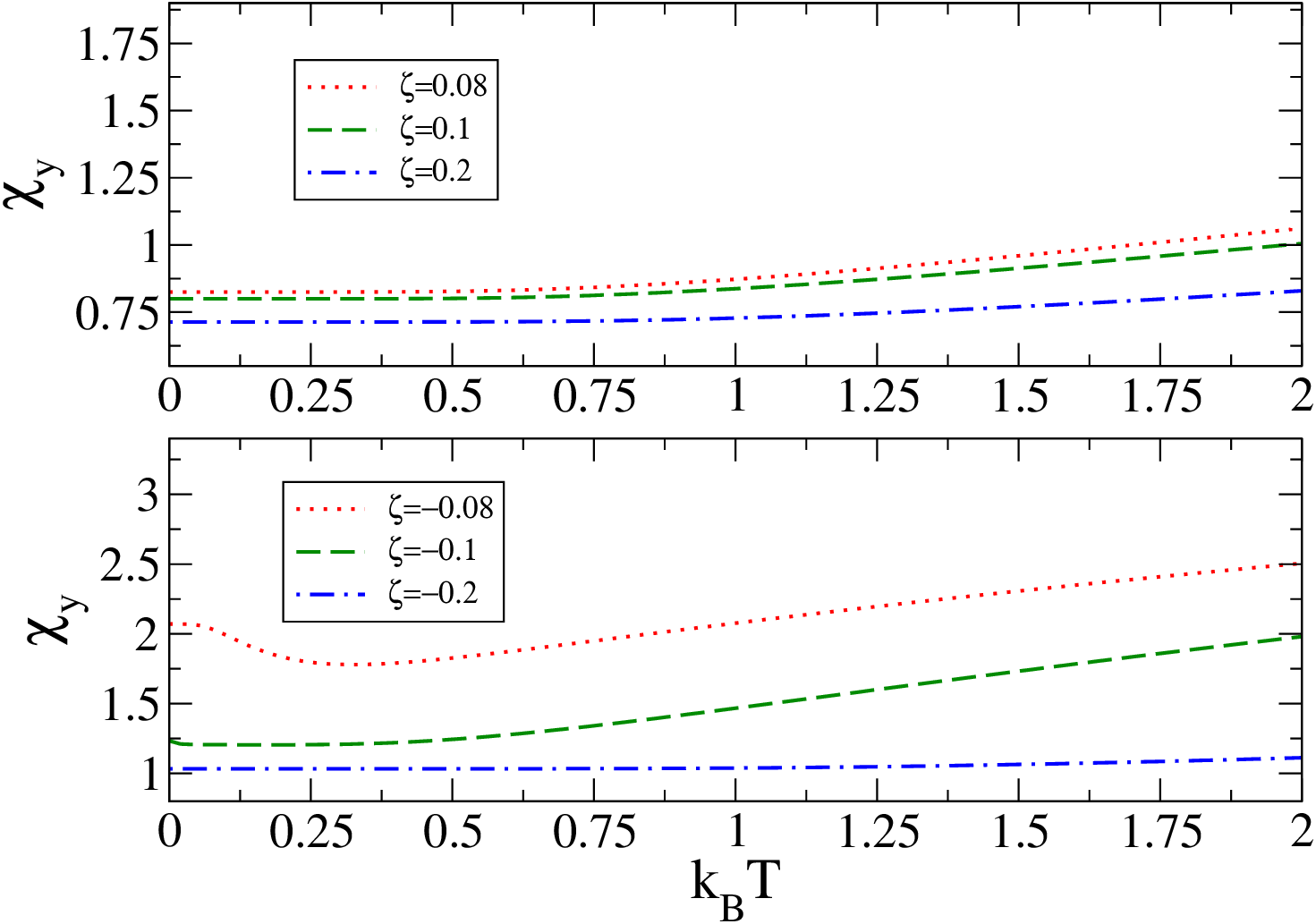,width=0.65\linewidth,clip=}
\caption{(Color online). The phase sensitivity indicator $\chi_{y}$  vs $k_B
T$. Upper panel: repulsive bosons, $\zeta >0$. Lower panel: attractive bosons, $\zeta<0$. Number of bosons $N=30$.
$k_BT$ is in units of $J$.}
\label{fig8}
\end{figure}
%Due to exchange between the ground and the highest excited states of the
%Hamiltonian (\ref{twomode}) under the transformation $U \rightarrow -U$ (see,
%e.g., \cite{diaz}), we shall perform our calculations in the attractive and repulsive regimes in correspondence to
%the same absolute value of $U$.
\begin{figure}[ht]
\epsfig{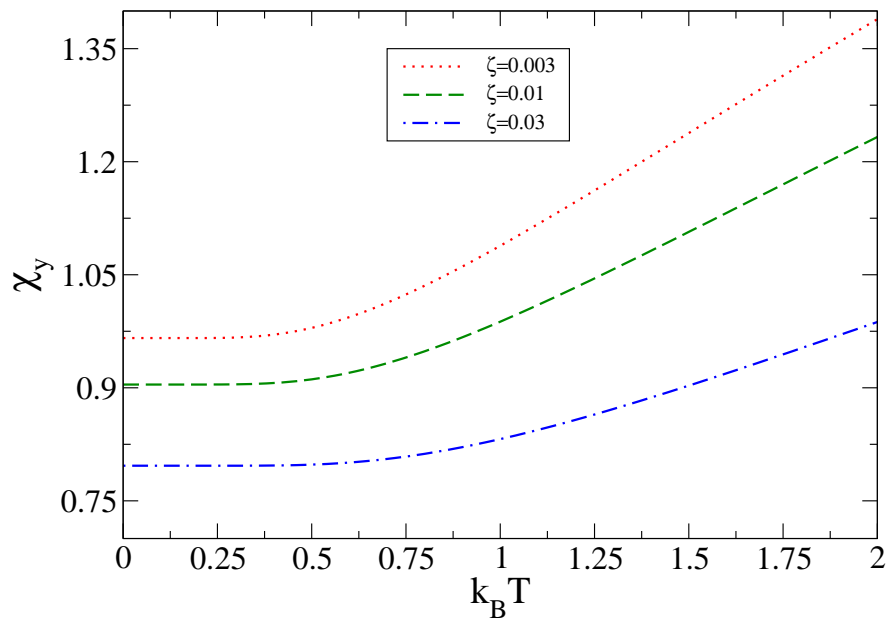}
\caption{(Color online). The phase sensitivity indicator $\chi_{y}$ vs $k_B
T$ for repulsive bosons, $\zeta >0$. Number of bosons $N=100$.
$k_BT$ is in units of $J$.}
\label{fig9}
\end{figure}

%The quantum Fisher informations $F_{QT,x}$ and $F_{QT,y}$,
%in terms of the coefficients of the expansion (\ref{eigenstate}),
%read
%\beqa
%\label{fiT2y}
%F_{QT,\gamma} &=&
%\sum_{j,k=0}^{N}\frac{(e^{-\beta E_j} -e^{-\beta E_k})^2}{2\,Z\,(e^{-\beta E_j}
%+e^{-\beta E_k})}\nonumber\\
%&&\big(\sum_{i=0}^{N}
%(c_{i}^{(j)}c_{i+1}^{(k)}
%\pm c_{i+1}^{(j)}c_{i}^{(k)})\,\sqrt{(i+1)(N-i)}\big)^{2} \;.\nonumber\\
%\eeqa
%In the above equation, the plus and minus signs hold for $F_{QT,x}$ and
%$F_{QT,y}$, respectively. Moreover, we observe that the terms with $i=N$
%are zero. Therefore the summation at the right hand side of Eq. (\ref{fiT2y}) is
%well defined even if $c_{N+1}^{(j,k)}$ are not defined in Eq. (\ref{eigenstate}).
%For $F_{QT,z}$, we have
%\beqa
%\label{ft2}
%F_{QT,z} &=&
%\sum_{j,k=0}^{N}\frac{(e^{-\beta E_j} -e^{-\beta E_k})^2}{2\,Z\,(e^{-\beta E_j}
%+e^{-\beta E_k})}\nonumber\\
%&&\big(\sum_{i=0}^{N}
%c_{i}^{(j)}c_{i}^{(k)}\,(2i-N)\big)^{2} \;.\nonumber\\
%\eeqa
By following the same path as for the pure states, the sufficient condition for the
multiparticle entanglement and the quantum metrology utility condition reads
\beq
\chi_{\gamma} <1
\;.\eeq

Let us focus on the usefulness of states (\ref{mixed}) for
interferometry based on rotations aroud the $\gamma$ axis.
%We consider the question whether the quantum metrology
%usefulness condition is more robust than that achieved for the transformation $\exp(i\theta\hat{S}_{z})$.
To calculate $\chi_{\gamma}$ and study them as functions of
temperature - for different interaction strength - we use Eq. (\ref{pezze}) in Eq. (\ref{chi}). In the absence
of boson-boson interaction, we have verified that $\chi_{\gamma} \ge 1$ for the investigated
temperatures. For nonzero interatomic interactions, $\chi_{x}$ as a function of the temperature
is shown in Fig. \ref{fig6} obtained for $N=30$. From these plots, we conclude that $\chi_{x}>1$ both for repulsive and attractive interactions.

The situation is different for $\chi_{y}$, Fig. \ref{fig8}.
The first very interesting thing is that in the repulsive regime, $\zeta>0$  - upper
panel - there are regions of temperature in which the condition $\chi_{y}<1$
holds. In particular, this condition is always met for $\zeta=0.2$, while for $\zeta=0.08$
and $\zeta=0.1$, to an increase of the temperature corresponds a progressive corruption of
$\chi_{y}<1$. The second important finding is that when $\zeta>0$, the robustness of the quantum
metrology usefulness against the temperature is enhanced by increasing the repulsion between
the bosons. In fact, for $\zeta=0.08$ and $\zeta=0.1$,
the temperatures of the above utility breakdown are,
respectively, $1.7016 J/k_B$ and $1.9724 J/k_B$.
The two aforementioned results hold also for an higher number of particles as shown in Fig. \ref{fig9} obtained with
$N=100$.  When $\zeta=0.003$ and $\zeta=0.01$, the two breakdown
temperatures are, respectively, to $0.63J/k_B$ and $1.04 J/k_B$.
For $\zeta=0.03$, the condition $\chi_y <1$ is always matched.
\begin{figure}[ht]
\epsfig{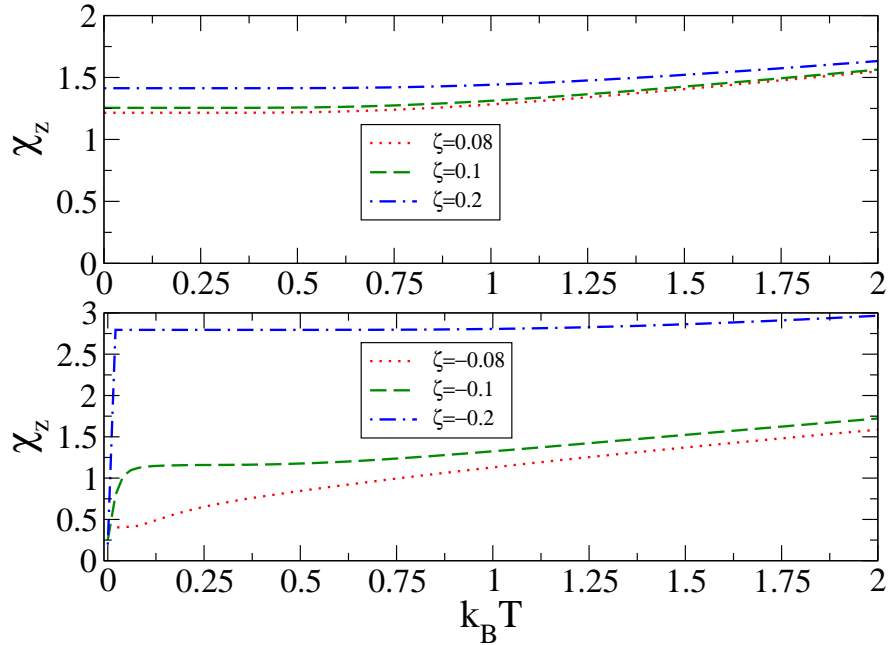}
\caption{(Color online). The phase sensitivity indicator $\chi_{z}$ vs $k_B
T$. Upper panel: repulsive bosons, $\zeta >0$. Lower panel:
attractive bosons, $\zeta<0$. Number of bosons $N=30$. $k_BT$ is in units of $J$.}
\label{fig4}
\end{figure}
Let us continue with $\chi_{z}$. In correspondence to finite values of $\zeta$, the results of our study are
reported in Fig. \ref{fig4}.  We can see that regions of
temperature exist in which $\chi_z<1$. This condition is met when the bosons
interact attractively, see the lower panel of Fig. \ref{fig4}. Again,
the thermal effects are those to cancel the quantum metrology usefulness.
For $\zeta=-0.08$ and $\zeta=-0.1$, the breakdown of quantum metrology
utility occurs, respectively, at $0.7595 J/k_B$ and $0.03603 J/k_B$.
For this kind of interferometry the stronger is the attraction between the bosons the smaller is the themal region in which sub shot-noise precisions can be achieved.
When $\zeta=-0.2$, the ground-state of the system is very close to a Schr\"{o}dinger cat state \cite{cats}.
From the lower panel of Fig. \ref{fig4}, it can be seen that the thermal softening of
the condition of $\chi_z<1$ strongly depends on the system ground-state.
In particular, the more this is different from a cat-like state the more is robust
the quantum metrology usefulness. 
%This may be related to the fact the smaller is $|\zeta|$ the greater is the thermal enhancement of the on-site number fluctuation \cite{inp}. 
From the upper panel of Fig. \ref{fig4}, we observe that in the repulsive regime there is no state useful for interferometry based on rotations about the $z$ axis.

We conclude this subsection by summarizing that in the attractive regime even in the presence of the temperature is possible to preserve the possibility to achieve sub-shot noise precisions by employing rotations around the $z$ axis. Within this kind of interferometry, to prepare the system as far as possible from Schr\"{o}dinger cat state is crucial to prevent the breakdown of the quantum metrology usefulness of a given thermal state. In the repulsive regime as well the thermal effects does not destroy the quantum metrology utility achieved at zero temperature for rotations about the $y$ axis. In this case, the more the ground-state is closer to the twin Fock state the wider is the thermal region where the quantum metrology usefulness condition is met. To support this consideration in a more complete way, in the next section we shall study the breakdown temperature as a function of the number of trapped bosons in correspondence to different interaction strengths.

\section{The breakdown temperature}

This section deals with the breakdown temperature of the quantum metrology usefulness. From our analysis we know $k_BT/J$ with $J$ given by Eq. (\ref{hopping}). The breakdown temperature thus will be known once solved the eigenvectors problem associated to $V_{DW}(x)$. Let us suppose that \cite{fermidyn}
\beq
V_{DW}(x)=A x^4+B(e^{-Cx^2}-1)
\eeq
and that the transverse trapping frequency $\omega_{\bot}= 2\pi \times 160$kHz.
This frequency, for instance, with $^{23}$Na atoms, corresponds to a transverse
confinement length $a_{\bot}=\sqrt{\hbar/(m\omega_{\bot})}=0.05 \mu$ m. By expressing the lengths in units of $a_{\bot}$ and the energies in units of
$\hbar \omega_{\bot}$, we choose $A=10^{-3}$, $B=1$, and $C=4$, that gives rise to $J=0.032$.  For the sensitivity
indicator $\chi_{y}$, the two breakdown temperatures with $N=30$  - upper panel of Fig. \ref{fig8} - are $416$nK and $483$nK, respectively. For $N=100$ - Fig. \ref{fig9} - the corresponding breakdown temperatures are $154$nK and $255$nK.  As for what concerns $\chi_z$, we have - lower panel of Fig. \ref{fig4} - $186$nK and $9$nK.
\begin{figure}[ht]
\epsfig{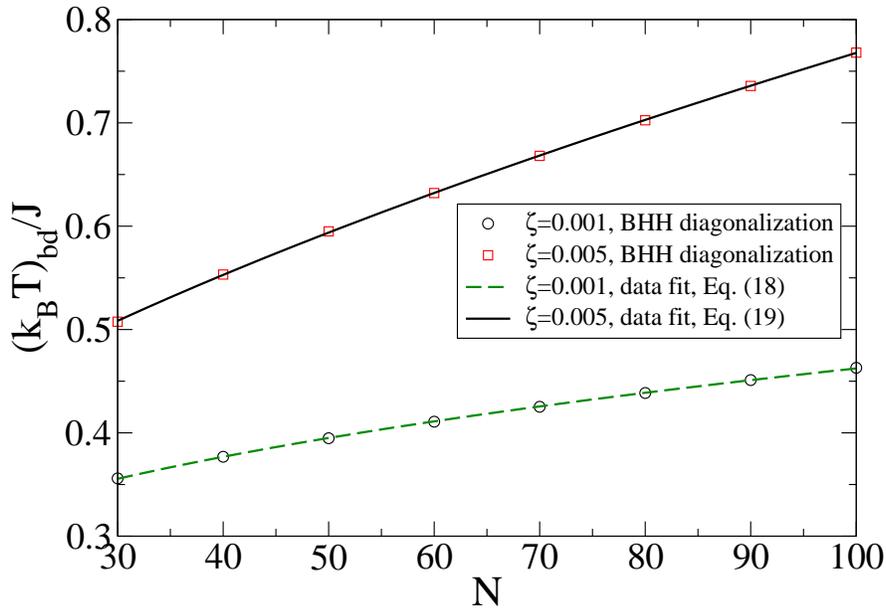}
\caption{(Color online). The breakdown scaled temperature $(k_BT)_{bd}/J$ vs $N$. Here BHH stays for Bose-Hubbard Hamiltonian.}
\label{fig10}
\end{figure}
Let us focus, now, on rotations about the $y$ axis .The behavior of the scaled breakdown (bd) temperature - $(k_BT)_{bd}/J$ - in terms of the size of the system is a very crucial issue. We report this study in Fig. \ref{fig10}. In this figure we have plotted $(k_BT)_{bd}/J$ as a function of the number of bosons $N$. We have carried out such an analysis for two different interaction strengths; $\zeta=0.001$ and $\zeta=0.005$. It can be observed that, once fixed the boson-boson interaction, the greater is $N$ the higher is the breakdown temperature. On the other hand, in correspondence to the same number of particles, to a stronger repulsion between the bosons corresponds an higher breakdown thermal energy. In order to get the empty circles,  $\zeta=0.001$,  and the empty squares, $\zeta=0.005$, we have followed the following procedure. We have fixed $N$ time by time and found the corresponding eigenvectors of the Hamiltonian (\ref{twomode}), after that we have calculated $\chi_y$ via Eq. (\ref{chi}) with $\gamma=y$, and estimated the scaled temperature at which $\chi_y=1$. Note that  - with the above choice of the parameters - when $\zeta=0.001$ the bd temperature for $N=100$ is $3.6\mu$k; when $\zeta=0.005$ the bd temperature for $N=100$ is $5.9\mu$k. We have, then, employed a fitting procedure to infer the scaling law of the breakdown temperature with $N$ from the diagonalization data got by the above described method. The dashed line - $\zeta=0.001$ - and the solid one - $\zeta=0.005$ - represent the curves achieved by fitting the diagonalization points. For $\zeta=0.001$ we have obtained
\beq
\label{bd1}
\frac{(k_B\,T)_{bd}}{J}=A_0+A_1 \,N^{1/4}\,\ln N
\;\eeq
with $A_0=0.226639$ and $A_1=0.0161871$, while for $\zeta=0.005$ the fitting procedure has provided
\beq
\label{bd2}
\frac{(k_B\,T)_{bd}}{J}=A_2+A_3\, N^{1/2}\,\ln N
\;\eeq
with $A_2=0.332222$ and $A_3=0.00945721$.
The scaling laws (\ref{bd1}) and (\ref{bd2}) show that the exponent of $N$ depends on the interaction strength.

\section{Conclusions}
We have considered a bosonic Josephson junction realized via a one-dimensional double-well trap
confining $N$ ultracold and dilute atoms. By employing the two-site Bose-Hubbard model as theoretical tool,
we have analyzed to which extent the junction states allow for achieving sub shot-noise precisions in phase shift estimation.
To do this, we have calculated the quantum Fisher
information and used it to study the phase sensitivity indicator. We have
employed this quantity to analyze the utility of the junction states for
the quantum metrology.

At zero temperature we have analyzed the phase sensitivity indicators as
functions of the boson-boson interaction by studying the influence of this interaction on the
utility of the atomic junction ground-states when they are used as input states
for atomic interferometry. The multiparticle entangled bosonic Josephson junction ground-states achievable with
attractive interactions provides sub-shot-noise precisions for interferometry based
on rotations about $z$ axis. When the bosons are repulsively interacting, bosonic Josephson junction ground-states are useful for
interferometry based on rotations about the $x$ and $y$ axes.

In the presence of the temperature we have studied the phase sensitivity indicators as functions of the temperature in correspondence to different interaction strengths.
We have pointed out that that even at finite temperature ranges of temperatures exist
over which the bosonic Josephson junction states are useful input states for quantum interferometry based on rotations about the $y$ and $z$ axes.
In the former case, we have found that an increase of the repulsion between the bosons enhances the robustness of the quantum metrology usefulness against the temperature; in the latter situation the robustness
enhancement takes place by lowering the absolute value of the interaction
strength. When the temperature is sufficiently high, the possibility to achieve sub shot-noise precisions is destroyed.
We have evaluated the temperature at which occurs the breakdown of the aforementioned usefulness. We have, moreover, analyzed the effects of the size of the system on the breakdown temperature by finding for a scaling law in terms of the number of bosons. We observed that the greater is the number of particles, the higher is the aforementioned temperature.\\

The present work has been supported by University of Padova, Progetto Giovani: "Many Body Quantum Physics and Quantum Control with Ultracold Atomic Gases" and partially by University of Padova, Progetto di Ateneo: "Quantum Information with Ultracold Atoms in Optical Lattices". GM thanks Giulia Ferrini, Lorenzo Maccone, Luca Pezz\'e, and Luca Salasnich for useful comments and suggestions.\\

\end{document}